\pdfoutput=1
\documentclass[letter]{aa}

\usepackage{ulem}
\usepackage{amsmath,amssymb}
\usepackage[varg]{txfonts}

\usepackage{natbib}

\usepackage{color}
\usepackage{graphicx}

\usepackage{epstopdf}

\usepackage{tablefootnote}

\usepackage{gensymb}
\usepackage{stfloats}

\usepackage[utf8]{inputenc}
\usepackage{hyperref}
\usepackage{booktabs,multirow}

\definecolor{mygray}{gray}{0.6}
\definecolor{darkblue}{rgb}{0.0, 0.0, 0.8}

\newcommand{\se}[1]{Sect.~\ref{sec:#1}}
\newcommand{\eq}[1]{Eq.~(\ref{eq:#1})}

\newcommand{\fg}[1]{Fig.~\ref{fig:#1}}
\newcommand{\Fg}[1]{Figure~\ref{fig:#1}}

\begin{document}

\title{What pebbles are made of: Interpretation of the V883 Ori disk}
\titlerunning{Interpretation of the V883 Ori snow line}
\author{Djoeke Schoonenberg\inst{1}, Satoshi Okuzumi\inst{2}, and Chris W. Ormel\inst{1}}
\institute{Anton Pannekoek Institute for Astronomy, University of Amsterdam, Science Park 904, 1090 GE Amsterdam, The Netherlands\label{inst1} \\
\email{d.schoonenberg@uva.nl}
\and
Department of Earth and Planetary Sciences, Tokyo Institute of Technology, Meguro, Tokyo, 152-8551, Japan\label{inst2}}
\date{\today}

\abstract{Recently, an Atacama Large Millimeter/submillimeter Array (ALMA) observation of the water snow line in the protoplanetary disk around the FU Orionis star V883 Ori was reported. The radial variation of the spectral index at mm-wavelengths around the snow line was interpreted as being due to a pileup of particles interior to the snow line. However, radial transport of solids in the outer disk operates on timescales much longer than the typical timescale of an FU Ori outburst ($10^{1}$--$10^{2}$ yr). Consequently, a steady-state pileup is unlikely. We argue that it is only necessary to consider water evaporation and re-coagulation of silicates to explain the recent ALMA observation of V883 Ori because these processes are short enough to have had their impact since the outburst. Our model requires the inner disk to have already been optically thick before the outburst, and our results suggest that the carbon content of pebbles is low.}

\maketitle

\section{Introduction}\label{sec:introduction}
Recently, \citet{2016Natur.535..258C} reported that the Atacama Large Millimeter/submillimeter Array (ALMA) has for the first time detected the water snow line in a circumstellar disk. The host of the disk, the T Tauri star V883 Ori, is undergoing a FU Orionis type outburst and is therefore very luminous \citep{1993ApJ...412L..63S,2001ApJS..134..115S,2014prpl.conf..387A}. The outburst has pushed the snow line out to a larger radial distance than in the quiescent phase, making it detectable with ALMA. Since the disk was observed at wavelengths of 1.289 mm and 1.375 mm at high angular resolution, it was possible to constrain the spectral index $\alpha$ as a function of distance from the star. It was also found that the inner disk (interior to the snow line) is optically thick, whereas the region outside the snow line is optically thin \citep{2016Natur.535..258C}. The observation was interpreted in terms of a pileup of particles interior to the snow line in a low-viscosity disk caused by a lower fragmentation threshold for dry (water-free) particles than for icy particles \citep{2015ApJ...815L..15B}.

In this Letter we argue that a solids pileup interior to the snow line is not likely to be the correct explanation for the ALMA observation of the snow line around V883 Ori because the average FU Ori outburst duration ($\sim$$10^{2}$ yr) is shorter than the pileup timescale ($\sim$$10^{4}$ yr; see \se{pile}). Therefore, post-outburst pileups have not yet materialized. In this Letter we present an alternative interpretation of the data presented by \citet{2016Natur.535..258C}, accounting only for water evaporation and re-coagulation of silicates. Our model is simple on purpose and we do not aim to fit the data perfectly. Rather, we aim to capture three main features of the ALMA observation:
\begin{enumerate}
\item The mm-emission is optically thick interior to the snow line.
\item There is a peak in the spectral index $\alpha$ of $\sim$$3.8$ just exterior to the snow line.
\item The spectral index $\alpha$ decreases toward $\sim$$3.5$ in the outer disk.
\end{enumerate}
In \se{model} we outline our disk model. In \se{results} we present the results of two models that match the criteria above as well as of three models that do not. We summarize our key findings and discuss possible improvements and avenues for future research in \se{discussion}.

\section{Model}\label{sec:model}
\begin{figure*}[t]
        \centering
                \includegraphics[width=0.33\textwidth]{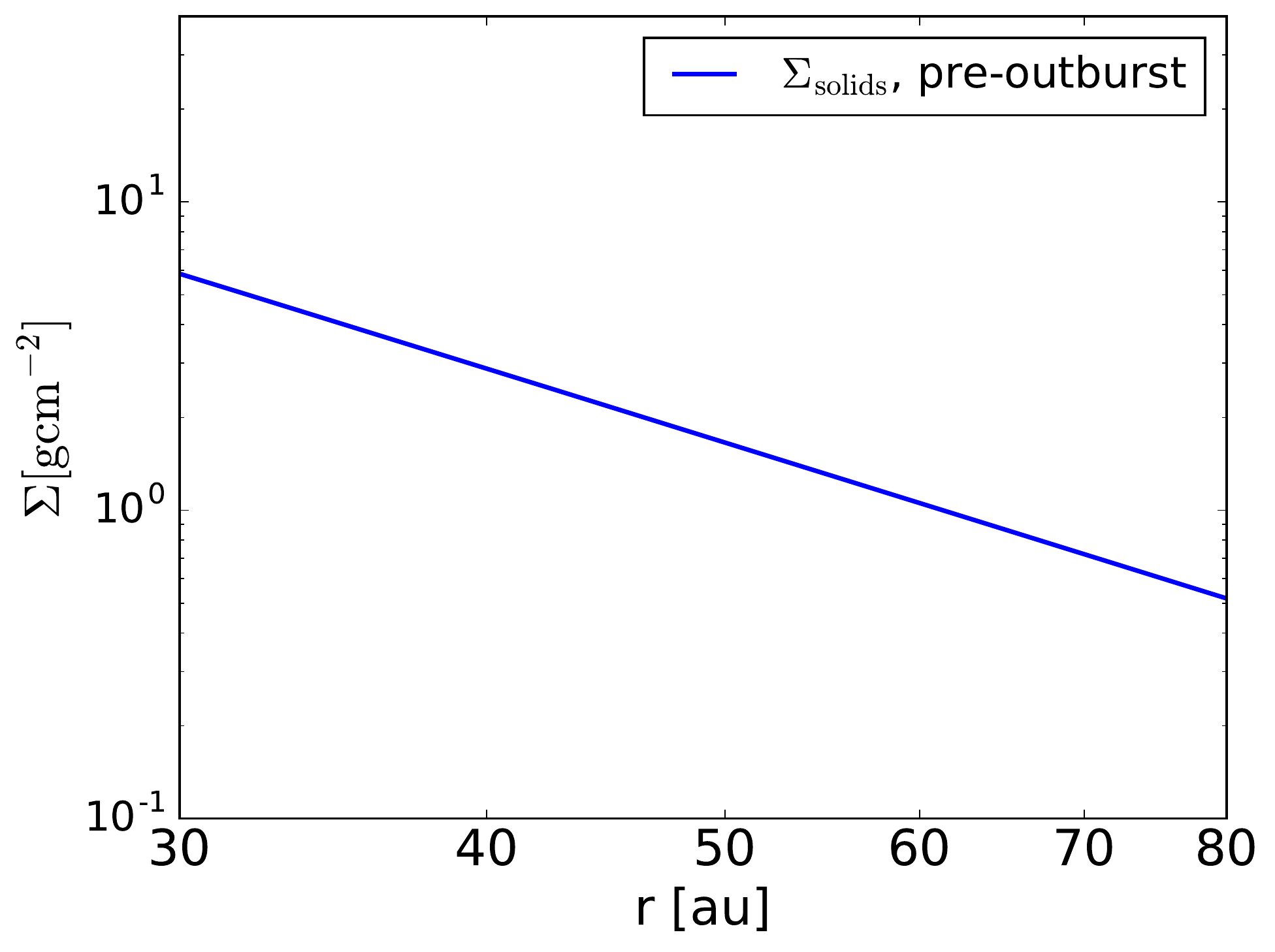}
                \includegraphics[width=0.33\textwidth]{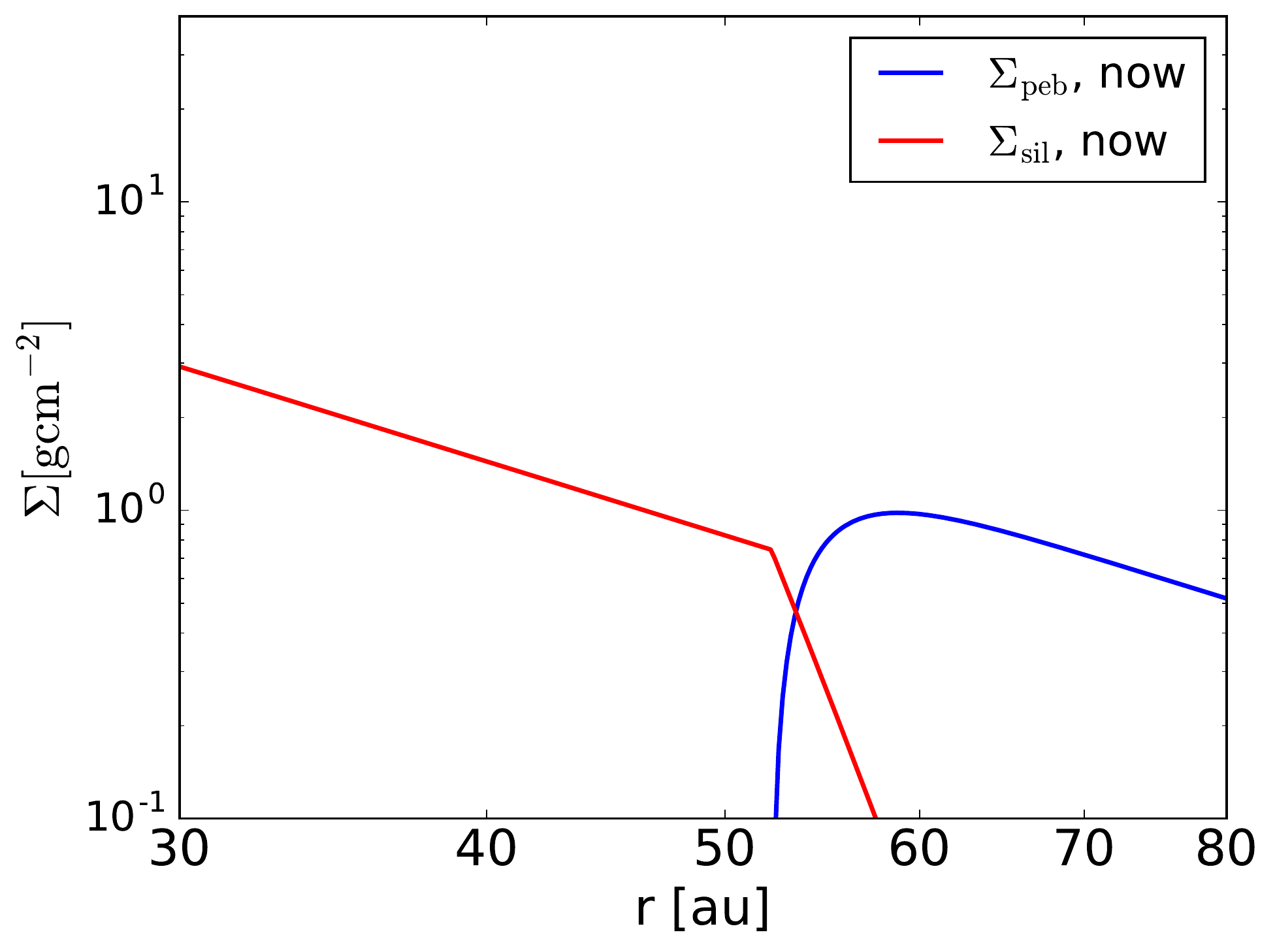}
                \includegraphics[width=0.33\textwidth]{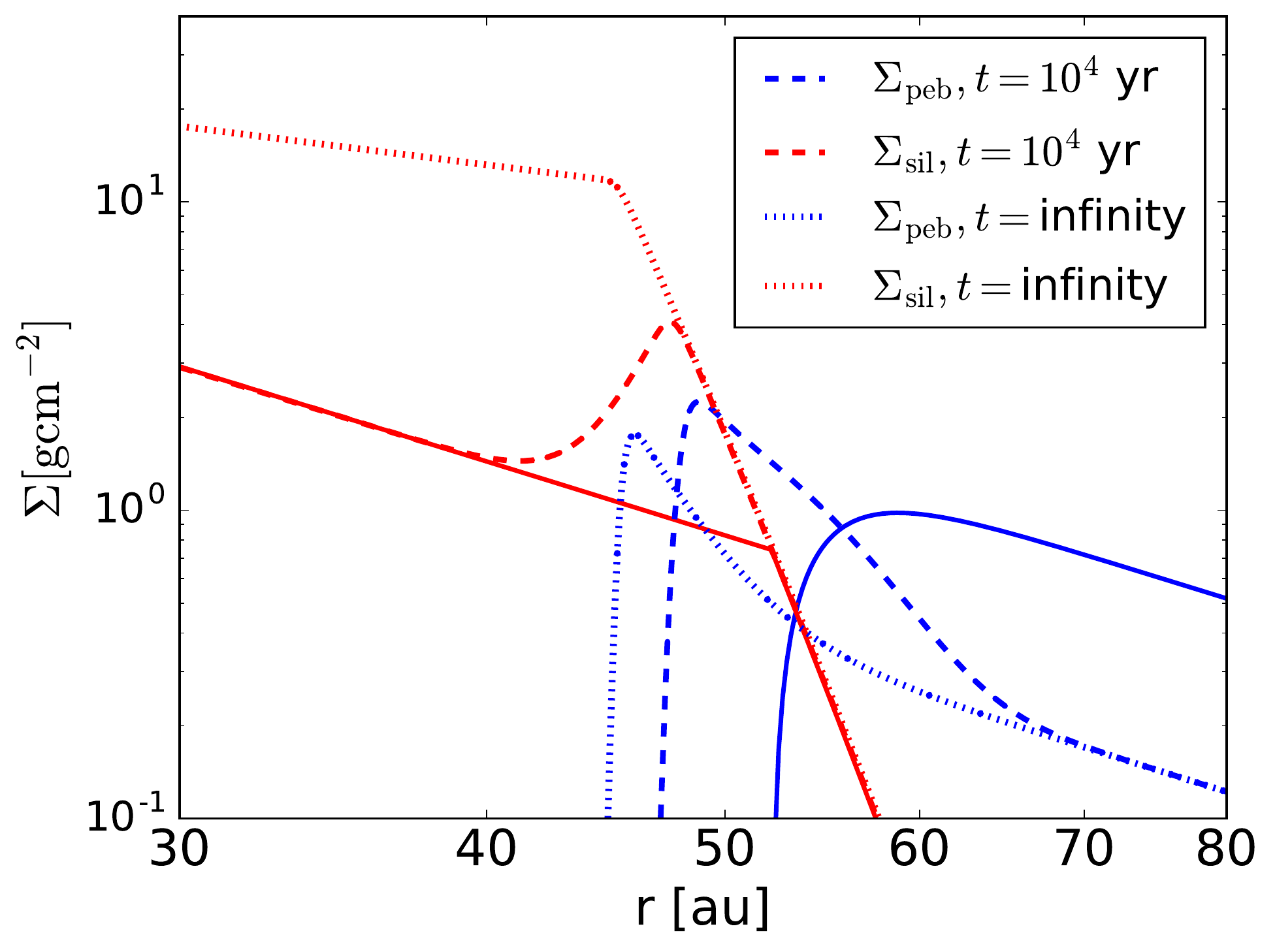}
        \caption{Solids surface density profiles $\Sigma$ in the V883 Ori disk at three different points in time. {\it Left}: Before the outburst, the snow line was located at around 2 au. In the domain plotted above the drifting pebbles consist of ices and silicates. {\it Middle}: During the outburst, the snow line has moved to $\sim$50 au, exterior to which the pebbles still contain water ice and silicates. The surface density profile of icy pebbles is given by the blue line. Interior to $\sim$$50$ au the pebbles have disintegrated and smaller silicate particles remain (red line). We expect that the V883 Ori disk is presently in this state. {\it Right}: Assuming that the disk temperature remains at $T_{\rm{post}}$ after the outburst, eventually a pileup of silicate particles interior to the snow line (red lines) is realized owing to their smaller drift velocity. There is also a pileup in the icy pebble surface density distribution (blue lines) owing to outward diffusion and re-condensation \citep{2017arXiv170202151S}. The dashed lines correspond to the situation after $10^{4}$ yr and the dotted lines correspond to the steady-state solution. The solid lines are the same as in the middle panel. Since the typical decay timescale of an FU Ori outburst is $\sim$$100$ yr, we do not expect to reach either of these states.\label{fig:sketch}}
\end{figure*}

\subsection{Disk model}
\label{sec:disk-model}
The gas surface density profile as a function of radial separation from the star $r$ is taken to be
\begin{equation}\label{eq:gas}
\Sigma_{\rm{gas}} = 50 \left(\frac{r}{40 \: \rm{au}}\right)^{-1.5} \rm{g} \: \rm{cm}^{-2}
,\end{equation}
which corresponds to seven times the minimum mass solar nebula and leads to disk masses in agreement with the results of \citet{2016Natur.535..258C}.
We adopted the following temperature profiles before and after the outburst (denoted by the subscripts `pre' and `post', respectively),
\begin{equation}\label{eq:temp}
T_{\rm{pre}} = 150 \left(\frac{r}{2.5 \: \rm{au}}\right)^{-0.5} \rm{K} ; \: \: \: \: \: {\it T}_{\rm{post}} = 150 \left(\frac{{\it r}}{45.3 \: \rm{au}}\right)^{-0.5} \rm{K},\end{equation}
where the power-law indices correspond to a passively irradiated disk \citep{1987ApJ...323..714K, 2016Natur.535..258C}. We estimated the thermal relaxation time to be \hbox{$\mathcal{O}$(1 yr)}, which is shorter than the outburst timescale and therefore we identify $T_{\rm{post}}$ with a thermally relaxed disk.

We assumed that the solids content of the disk is made up of pebbles, which are characterized by a typical size \citep{2012A&A...539A.148B,2016A&A...586A..20K, SatoEtal2016}. Before the outburst, the pebble surface density $\Sigma_{\rm{peb}}$ is determined by a constant pebble mass flux $\dot{M}_{\rm{peb}}$ and drift velocity $v_{\rm{drift}}$,\begin{equation}
\Sigma_{\rm{peb}} = \frac{\dot{M}_{\rm{peb}}}{2 \pi r v_{\rm{drift}}} ; \: \: \: \: \: v_{\rm{drift}} = \frac{v_{\rm{gas}} + 2 \eta v_{K} \tau_{\rm{peb}}}{1 + \tau_{\rm{peb}}^{2}}
,\end{equation}
where $v_{\rm{gas}}$ is the radial velocity of the gas and $\eta v_{K}$ is the deviation of the azimuthal gas speed from the Keplerian velocity $v_{K}$ \citep{Weidenschilling1977,NakagawaEtal1986}. In the Epstein drag regime, the dimensionless stopping time at the disk midplane $\tau_{\rm{peb}}$ is equal to \citep{2012A&A...539A.148B}
\begin{equation}\label{eq:stoppingtime}
\tau_{\rm{peb}} = \frac{\pi}{2} \frac{\rho_{\bullet} a_{\rm{peb}}}{\Sigma_{\rm{gas}}}
,\end{equation}
where $\rho_{\bullet} = 1.5 \: \rm{g} \: \rm{cm}^{-3}$ is the internal pebble density and $a_{\rm{peb}}$ is the pebble radius. The pre-outburst solids surface density is shown in the left panel of \fg{sketch}.

We tuned the value of $\dot{M}_{\rm{peb}}$ to get the best result for a particular model (\se{results}). The actual value of $\dot{M}_{\rm{peb}}$ and the pre-factors in $\Sigma_{\rm{gas}}$ (\eq{gas}), $T_{\rm{pre}}$, and $T_{\rm{post}}$ (\eq{temp}) are not very meaningful in this work since our results are degenerate between these quantities: from Eqs. (3)--(5) one can show that $\Sigma_{\rm{peb}} \propto \dot{M}_{\rm{peb}} \Sigma_{\rm{gas}}$ for pebble-sized particles that have $v_{\rm{drift}} \propto \tau_{\rm{peb}}$, and the location of the post-outburst snow line depends both on $T_{\rm{post}}$ and on $\dot{M}_{\rm{peb}}$ \citep{2017arXiv170202151S}.

\subsection{Evaporation and condensation}
After the onset of the outburst the disk heats up and the icy pebbles interior to the new snow line location evaporate, resulting in a post-outburst solids surface density distribution that is sketched in the middle panel of \fg{sketch}. We adopted the `many-seeds' model of \citet{2017arXiv170202151S}, in which icy pebbles beyond the snow line consist of many micron-sized silicate particles that are `glued' together by water ice. When icy pebbles evaporate, micron-sized bare silicate particles are left behind, and these particles subsequently re-coagulate (\se{pebblecom}).

We assumed that the silicate surface density profile $\Sigma_{\rm{sil}}$ closely follows the equilibrium (saturated) water vapor surface density profile $\Sigma_{\rm{vap, sat}}$, which is obtained from the Clausius-Clapeyron equation \citep{2017arXiv170202151S}
\begin{equation}
\Sigma_{\rm{sil}} = \rm{min} \left(\frac{{\it f}_{\rm{sil}}}{1 - {\it f}_{\rm{sil}}} \Sigma_{\rm{vap, sat}}, \: {\it f}_{\rm{sil}} \Sigma_{\rm{peb}}\right)
,\end{equation}
where we take the silicate fraction $f_{\rm{sil}}$ of icy pebbles beyond the snow line equal to 0.5 \citep{2003ApJ...591.1220L}.

\subsection{Re-coagulation of silicates}\label{sec:pebblecom}
The icy pebbles are vertically settled and just after the evaporation of their hosts, the silicate particles are as well \citep{2016A&A...596L...3I}. The vertical diffusion timescale for the released silicate particles is given by
\begin{equation}
t_{\rm{diff}} = \frac{H_{\rm{gas}}^{2}}{\nu} = \frac{1}{\alpha_{T} \Omega} \approx 2.3 \times 10^{4} \left(\frac{\alpha_{T}}{10^{-3}}\right)^{-1} \left(\frac{r}{30 \: \rm{au}}\right)^{1.5} \: \rm{yr}
,\end{equation}
where we have taken the turbulent diffusivity equal to the viscosity $\nu = \alpha_{T} H_{\rm{gas}}^{2} \Omega$ \citep{1973A&A....24..337S} with $\alpha_T = 10^{-3}$ throughout this work and $H_{\rm{gas}}$ the gas scale height. The decay timescale of an FU Ori outburst $t_{\rm{outburst}}$ is typically on the order of decades or centuries \citep{1990IAUS..137..229R,1996ARA&A..34..207H,2000AIPC..522..411K}, which was also found for V883 Ori specifically \citep{1993ApJ...412L..63S}. Because $t_{\rm{diff}} \gg t_{\rm{outburst}}$, we concluded that the silicates are currently still residing in the settled layer, which speeds up coagulation. For our benchmark model we find coagulation timescales $t_{\rm{coag}} \sim \mathcal{O}(10 \: \rm{yr})$ at the post-outburst snow line location.

Since the coagulation timescale is shorter than the outburst timescale, the silicate particles have had time to settle into a coagulation-fragmentation equilibrium \citep{2011A&A...525A..11B}. The maximum silicate particle size $a_{\rm{max}}$ is determined by the fragmentation threshold velocity, which for silicates is on the order of 1 $\rm{m}\:\rm{s}^{-1}$ \citep{2010A&A...513A..56G,2010A&A...513A..57Z}.
Interior to the snow line, turbulent relative velocities are of this order for a particle size \hbox{$a_p \approx 300 \: \mu$m}; therefore, we take $a_{\rm{max}} = 300 \: \mu$m. For simplicity we adopted a standard Mathis-Rumpl-Nordsieck (MRN) index of $-3.5$ \citep{1977ApJ...217..425M}.

\begin{figure*}[t]
        \centering
                \includegraphics[width=0.49\textwidth]{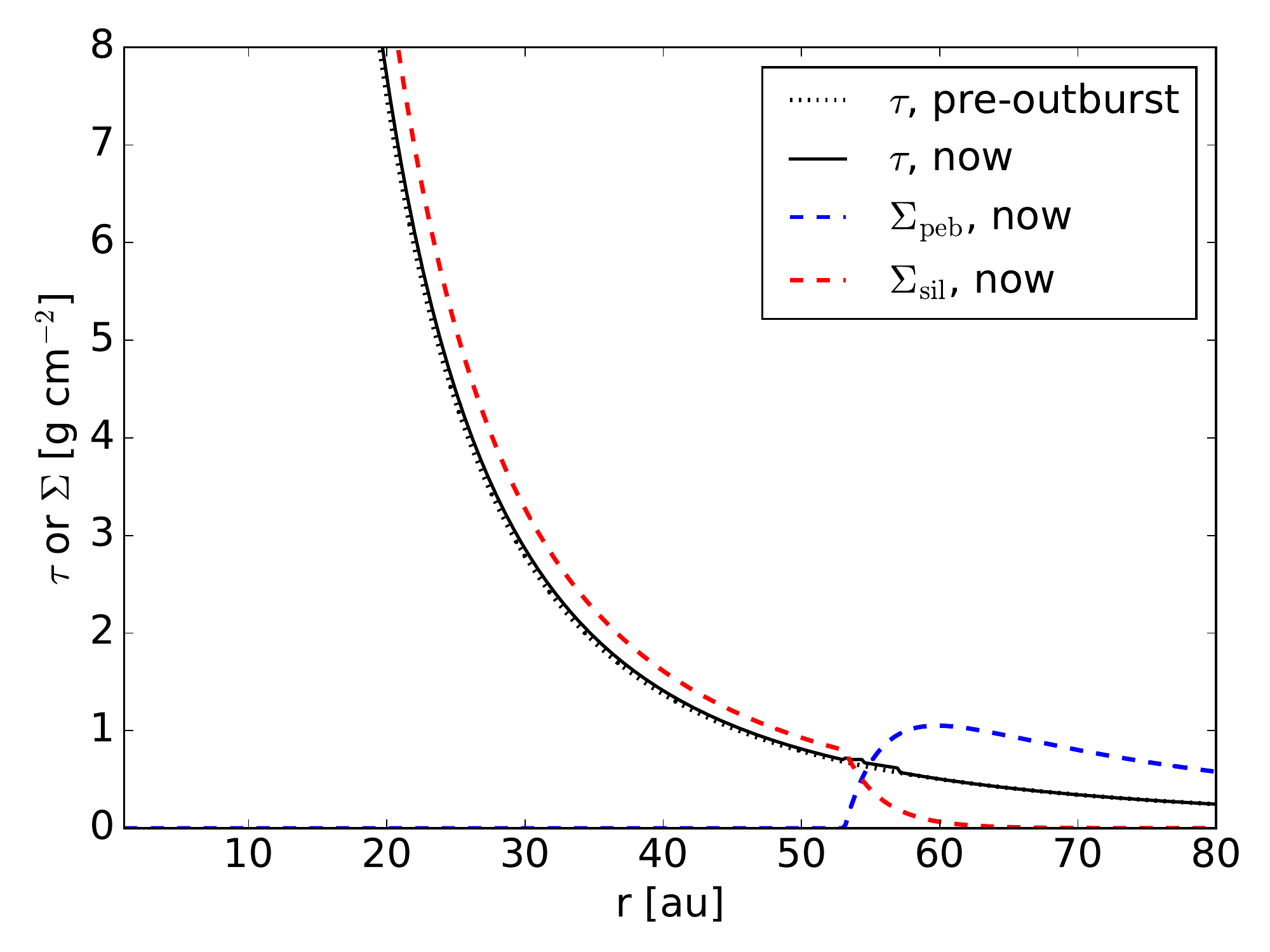}
                \includegraphics[width=0.49\textwidth]{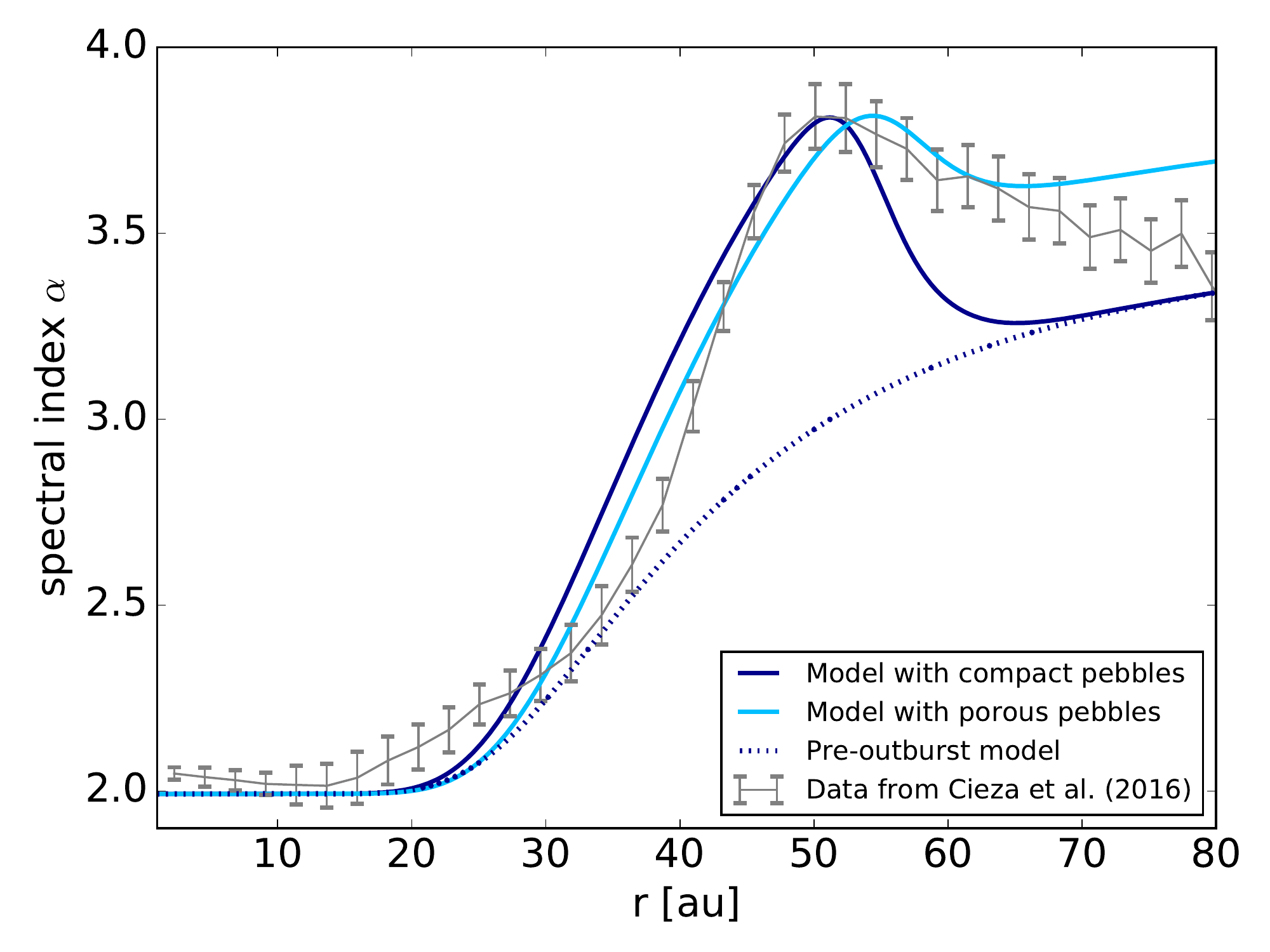}
        \caption{{\it Left}: Optical depth $\tau$ and surface density profiles $\Sigma$ as a function of distance from the star for compact pebbles. {\it Right}:  Spectral index as a function of distance from the star for our benchmark model with compact pebbles of initial size 0.5 cm pre-outburst (dotted dark blue line) and post-outburst (solid dark blue line), and for our post-outburst model with porous pebbles of packing fraction 0.1 and initial size 5.0 cm (light blue line). The gray line with error bars corresponds to the data from \citet{2016Natur.535..258C}.\label{fig:results}}
\end{figure*}

\subsection{Pileup}\label{sec:pile}
If the outburst lasted long enough, eventually a pileup of solids interior to the snow line would occur because the silicate particles have a smaller drift velocity than the icy pebbles outside the snow line \citep{2015ApJ...815L..15B}. In the right panel of \fg{sketch} we show the situation after $10^{4}$ yr as well as the steady state. It can be seen that even after $10^{4}$ yr, the pileup has not yet spread throughout the inner disk. Because the outburst timescale is much shorter, we do not expect a significant pileup to ever be reached, and conclude that the V883 Ori disk is currently in the phase depicted in the middle panel of \fg{sketch}.

\section{Comparison with the ALMA observation}\label{sec:results}

In order to be able to compare our model results to the ALMA observation presented in \citet{2016Natur.535..258C}, we defined the spectral index as $\alpha \equiv \ln{(I_{218.0 \rm{GHz}} / I_{232.6 \rm{GHz}})} / \ln{(232.6 / 218.0)}$. The value $I_{\nu}$ is the intensity at frequency $\nu$, given by
\begin{equation}
I_{\nu} = B_{\nu} (1 - \exp{[-\tau_{\nu}}])
,\end{equation}
where $B_{\nu} = B_{\nu} (T_{\rm{post}})$ is the Planck function and $\tau_{\nu}$ is the optical depth along our line of sight, given by
\begin{equation}
(\cos i) \tau_{\nu}= \kappa_{\nu, \rm{sil}} \Sigma_{\rm{sil}} + \kappa_{\nu, \rm{peb}} \Sigma_{\rm{peb}}
,\end{equation}
where $i = 38.3 \degree$ is the inclination of the V883 Ori disk and $\kappa_{\nu, \rm{sil}}$ and $\kappa_{\nu, \rm{peb}}$ are the absorption opacities at frequency $\nu$ of the silicate particles and icy pebbles, respectively. We used the DIANA Opacity Tool\footnote{Publicly available at \url{http://dianaproject.wp.st-andrews.ac.uk/data-results-downloads/fortran-package/}} to calculate the absorption opacities for different particle sizes and compositions \citep{WoitkeEtal, ToonAckerman, MinEtal, DorschnerEtal, ZubkoEtal}.

Before calculating the spectral index $\alpha$, we smoothed the intensities with a (Gaussian) beam size of 12 au, corresponding to the 0.03 arcsec resolution reported in \citet{2016Natur.535..258C}.

\subsection{Benchmark model}
In our benchmark model the initial physical size of icy pebbles is constant throughout the disk. This leads to an initial pebble surface density profile proportional to $r^{-5/2}$ (\eq{stoppingtime}). The icy pebbles have an initial size of 0.5 cm (corresponding to a stopping time of $\sim$$0.03$ at 50 au) and zero porosity.
The benchmark results are shown in \fg{results}. In the left panel we show the present surface density of icy pebbles ($\Sigma_{\rm{peb}}$) and silicate particles ($\Sigma_{\rm{sil}}$). We also show the optical depth $\tau$ at 1.375 mm before and after the outburst.

In the right panel, we compare our benchmark model predictions for the variation of the spectral index $\alpha$ with the ALMA observation. All three criteria defined in \se{introduction} are met by the post-outburst benchmark model. The first criterion --- an optically thick inner disk --- is also met by the pre-outburst model, as reflected by $\alpha \rightarrow 2$ in the inner disk.

\subsection{Effect of porosity}
Increasing the porosity of the icy pebbles while increasing their physical size by the same factor does not change the results much because the stopping time of a pebble (in the Epstein regime) depends on the product of its filling factor and physical size (\eq{stoppingtime}). The only difference between a high porosity and large size model and a low porosity and small size model is that porous pebbles have a slightly higher opacity index $\beta$ than compact pebbles of the same stopping time \citep{2014A&A...568A..42K}. The light blue line in the right panel of \fg{results} shows our results for 5 cm pebbles with a packing fraction of 0.1. The fact that the data lie between the curves for the porous and compact pebble models suggests that the spectral index in the outer disk can be explained by a combination of the porous and compact models; for example, just outside the snow line pebbles have already grown to porous aggregates, while even further out they are still smaller and compact.

\subsection{Carbonaceous pebbles}
The match between the model predictions and the data becomes worse when the pebbles contain more carbonaceous grains. This is because the opacity index $\beta$ at millimeter wavelengths of pebbles decreases with increasing carbon content, leading to a lower spectral index in the optically thin region. The solid blue line in \fg{failed} gives the spectral index when 10\% of the silicate fraction of pebbles is substituted with carbonaceous materials. 

\subsection{Constant dimensionless stopping time}
In a drift-limited solids distribution, the dimensionless stopping time $\tau_{\rm{peb}}$ tends to be nearly constant throughout the disk \citep{2014A&A...572A.107L}. If we keep $\tau_{\rm{peb}}$ constant, the surface density profile of solids is proportional to $r^{-1}$, in contrast to $\propto r^{-5/2}$ for constant pebble size $a_{\rm{peb}}$ (benchmark model; \fg{results}). This leads to an optical depth profile that is too shallow compared to the optical depth profile observed by \citet{2016Natur.535..258C}. The results for our model with constant pebble stopping time are shown by the dashed blue line in \fg{failed}.

\subsection{Model without re-coagulation of silicates}
The dotted blue line in \Fg{failed} shows the result of our model without re-coagulation of silicate particles (\se{pebblecom}), demonstrating that our model does not match the data well if micron-sized silicate grains do not re-coagulate to larger sizes. In our benchmark model, the re-coagulated silicate grains follow a size distribution with $a_{\rm{max}} = 300 \: \mu$m, which corresponds to an optical size $x_{\rm{max}} = 2 \pi a_{\rm{max}} / \lambda \sim 1$. In the opacity model used in this work, the opacity and opacity index ($\beta$) of particles of optical size $\sim 1$ are a factor several larger at mm-wavelengths than those of much smaller or larger particles. Therefore, for $a_{\rm{max}} \sim 300 \: \mu$m, it is possible to have an optically thick inner disk while still having $\alpha \sim 3.8$ beyond the snow line, whereas for \hbox{1 $\mu m$} grains these two observed features cannot be reproduced simultaneously.

\begin{figure}[t]
        \centering
                \includegraphics[width=0.5\textwidth]{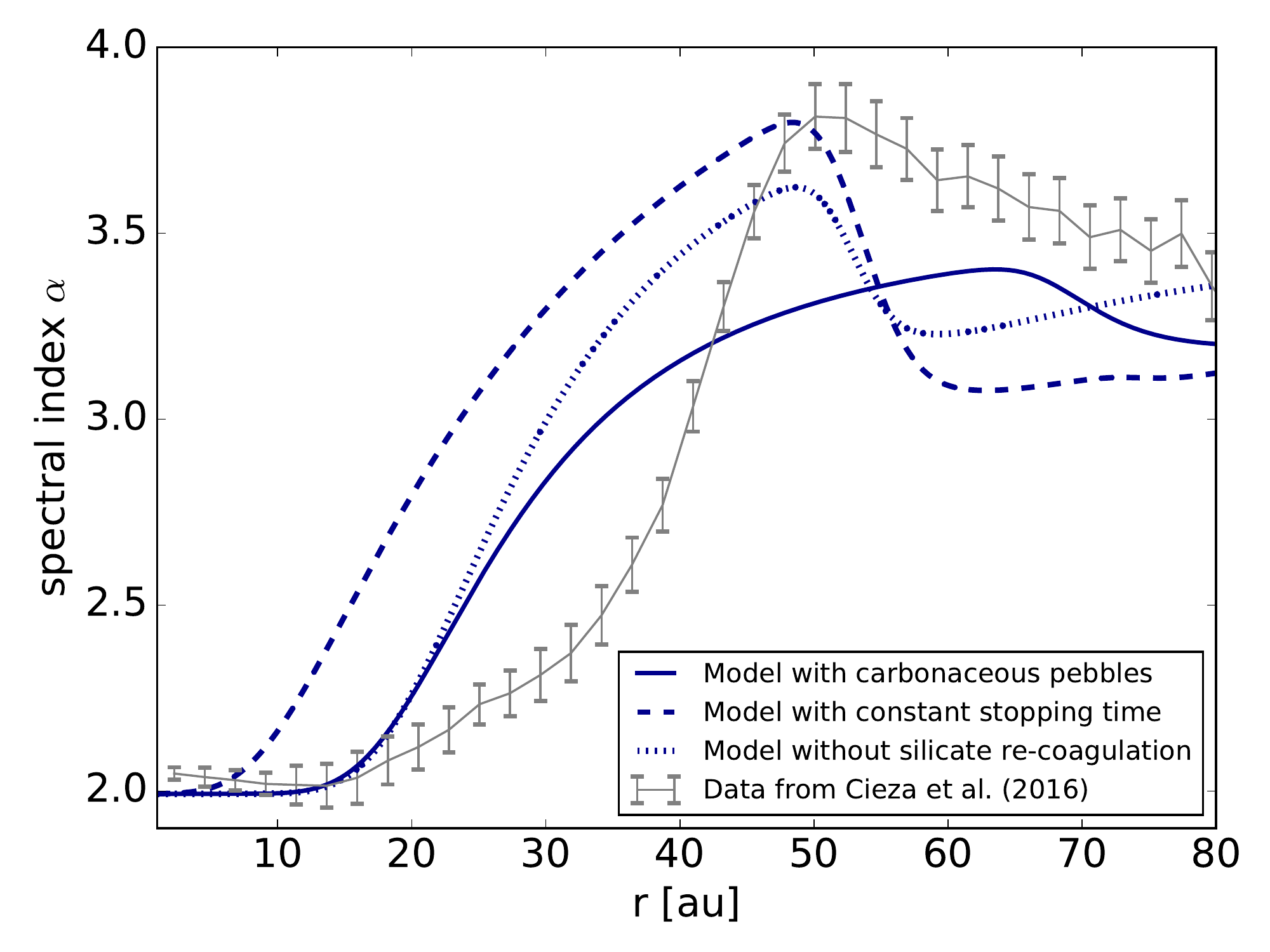}
\caption{Same as the right panel of \fg{results}, but for pebbles with 10\% carbon (solid blue line), for a model with constant stopping time throughout the disk (dashed blue line), and for a model with micron-sized-only silicate particles (dotted blue line).\label{fig:failed}}
\end{figure}

\section{Conclusions and discussion}\label{sec:discussion}
Our key findings can be summarized as follows:
\begin{itemize}
\item A simple model including only water evaporation and re-coagulation of silicates after an FU Ori outburst captures the characteristics of the ALMA observation of V883 Ori reported in \citet{2016Natur.535..258C}.
\item The model requires the inner disk to be already optically thick before the outburst.
\item Our model reproduces the ALMA observation best when we assume carbon-poor icy pebbles, which re-coagulate to $\sim$$300 \:\mu$m after evaporation.
\end{itemize}

We also found that an initially constant pebble size works better than an initially constant dimensionless stopping time because the former leads to a steeper gradient in the optical depth that better matches the observation (\fg{results}). A constant pebble size could be justified by a material property; for example, icy pebbles cannot grow beyond a certain size owing to a bouncing barrier, possibly induced by sintering \citep{2017arXiv170504778S}. Alternatively, a steep optical depth profile can be realized by adjusting the semimajor axis dependency of other disk quantities.

Naturally, the simplicity of our model implies several improvements. Firstly, the success of our model relies on the re-coagulation of $\mu$m silicate grains to $\sim$$300 \: \mu$m grains. A more sophisticated model would take into account radial variations in the maximum silicate particle size. Another possibility is that silicate seeds that are encapsulated in icy pebbles are not all micron-sized as assumed in this work, but already follow a size distribution.
Secondly, our pre-outburst pebble surface density profile does not take into account the $\rm{CO}_{2}$ and CO ice lines, which would both be located within the inner 100 au \citep{2011ApJ...743L..16O} and could lead to discontinuities in the initial pebble surface density profile, although this would probably be a minor effect \citep{2017A&A...600A.140S}.
However, we stress that the goal of this Letter is to offer an alternative and more physical explanation for the observation of V883 Ori than that presented in \citet{2016Natur.535..258C}, rather than a thorough fitting of our model to the data. If ALMA were to observe water snow lines in more disks around outbursting stars --- a phenomenon which is suspected to be very common for young stars \citep{1996ARA&A..34..207H,2000AIPC..522..411K,2014prpl.conf..387A} --- a dedicated parameter study could be a promising method to constrain the physical properties of pebbles (although parameters other than pebble size and composition have an effect on the opacity index as well \citep{WoitkeEtal}).

In this Letter we have neglected transport processes because the corresponding timescales are much longer than those of evaporation and re-coagulation. However, including transport processes would result in small temporal variations in the solids distribution after the outburst, which might be detectable by observing a disk around a young FU Ori object at different points in time. Also, it was recently proposed that the cooling down of a disk after an outburst could facilitate planetesimal formation by preferential re-condensation \citep{2017ApJ...840L...5H}, making FU Ori objects even more interesting to study from a planet formation perspective.

\begin{acknowledgements}
    D.S.\ and C.W.O\ are supported by the Netherlands Organization for Scientific Research (NWO; VIDI project 639.042.422). S.O. is supported by JSPS Grants-in-Aid for Scientific Research (No. 16K17661). We are thankful to the Earth and Life Science Institute (ELSI) in Tokyo for the support that made this collaboration possible. We would like to thank Michiel Min for providing a custom version of the DIANA Opacity Tool, Lucas Cieza for providing their data, and Tomas Stolker for a helpful discussion.
\end{acknowledgements}

\bibliography{FUoribib}

\end{document}